# Electric field assisted alignment of monoatomic carbon chains

Stella Kutrovskaya,[1,2,3] Igor Chestnov,[1,2,3] Anton Osipov,[3,4] Vlad Samyshkin,[3] Irina Sapegina,[5] Alexey Kavokin,[1,2,6,7] and Alexey Kucherik[3]

[1] School of Science, Westlake University, 18 Shilongshan Road, Hangzhou 310024, Zhejiang Province, China

[2]Institute of Natural Sciences, Westlake Institute for Advanced Study, 18 Shilongshan Road, Hangzhou 310024, Zhejiang Province, China

[3]Department of Physics and Applied Mathematics, Stoletov Vladimir State University, 600000 Gorkii street, Vladimir, Russia

[4]ILIT RAS — Branch of FSRC "Crystallography and Photonics" RAS, 1 Svyatoozerskaya, Shatura, 140700, Moscow region, Russia

[5] Udmurt Federal research center of the Ural Branch RAS, 34 Baramzina's street, 426067, Izhevsk, Russia

[6]Physics and Astronomy, University of Southampton, Highfield, Southampton, SO171BJ, United Kingdom

[7]Spin Optics Laboratory, St. Petersburg State University, 198504, Ulianovskaya str. 1, St. Petersburg, Russia

We stabilize monoatomic carbon chains in water by attaching them to gold nanoparticles (NPs) by means of the laser ablation process. Resulting nanoobjects represent pairs of NPs connected by multiple straight carbon chains of several nanometer lengths. If NPs at the opposite ends of a chain differ in size, the structure acquires a dipole moment due to the difference in work functions of the two NPs. We take advantage of the dipole polarisation of carbon chains for ordering them by the external electric field. We deposit them on a glass substrate by the sputtering method in the presence of static electric fields of magnitudes up to $10^5$ V/m. The formation of one-dimensional carbyne quasi-crystals deposited on a substrate is evidenced by high-resolution TEM and X-ray diffraction measurements. The original kinetic model describing the dynamics of ballistically flowing nano-dipoles reproduces the experimental diagram of orientation of the deposited chains.

## INTRODUCTION

Carbon-based materials such as graphene, fullerenes and carbon nanotubes provide a versatile platform for modern science and technology [1]. In this context, carbyne occupies an exceptional place among low-dimensional carbon allotropes. Carbyne is a monocrystal that consists of a linear chain of $sp$-hybridized carbon atoms. It may be considered as an ultimate 1D crystal. Two allotropes of carbyne: polyyne and cumulene are exceptionally interesting from the fundamental point of view and highly promising for applications in nano-electronic devices. They possess fascinating mechanical properties such as an outstanding elastic mod ulus [2]. Besides, carbyne has been regarded as an ideal (the thinnest possible) conductor with tunable electrical properties [3, 4]. Being an ideal 1D material, carbyne represents a multi-purpose platform for studies of electron, phonon and spin transport in 1D systems [1, 5].

Although, the existence of a sp-carbon allotrope was predicted a long time ago [6], the study of carbyne remained exclusively theoretical until recently. The reasons are the anticipated instability of an infinite 1D crystals of sp-hybridized carbon and their tendency to transform into other carbon allotropes at finite temperatures. The pioneering works devoted to the experimental investigation of carbyne [7, 8] evidenced the presence of few-atom segments of isolated linear wires. The recent breakthrough in the fabrication of carbyne [9] extended the limiting length of a single wire up to thousands of atoms by encapsulating it within a double-wall nanotube. Although the lengths of the chains realized in that study were remarkably large, it did not give access to the optical and electronic properties of the free-standing carbyne as the vicinity of a carbon nanotube heavily affects the properties of the encapsulated carbyne.

An alternative strategy to the fabrication of stable monoatomic carbon chains implies their capping with various end-groups [10] such as $sp^2$ groups terminating the carbon chains, transition metals, etc. A significant progress was recently achieved with the use of noble metal nanoparticles (NPs) as stabilizing agents. In particular, it was demonstrated [11] that the gold NPs of a few nanometer diameter are able to preserve carbyne chains from folding and bending even if the chain length exceeds tens of nanometers. Acting as heavy anchors, NPs can be used for the chain manipulation, in particular for its moving and stretching. Besides, NPs may be used as nanojunctions in electronic applications. Here we report on the new approach to the substrate deposition of carbyne wires, which enables for the spatial ordering of multiple chains stabilized by gold NPs. In contrast to previous studies, which were predominantly focused on the fabrication and investigation of individual carbyne wires, our method is highly convenient for the manipulation of a macroscopically large number of carbyne chains. The surface density of carbyne-NP complexes achieves 100 µm$^{-2}$ of in our experiments. We explore the sensitivity of the carbyne-NPs complexes to the electric field in order to arrange multiple carbyne wires parallel to each other. The present method enables obtaining bundles of tightly packed wires rather than individual carbyne chains, because the Van der Waals attraction between parallel chains favours formation of honeycomb-like quasi-crystal structures based on carbynes. The ensembles of aligned linear carbon chains obtained in our experiments may be employed for the realization of novel quantum nanoelectronic and nano-photonic devices including nanometer-size diodes and transistors, single and entangled photon emitters.

**RESULTS**

**The experimental approach to the growth and manipulation of monoatomic carbon chains**

Stable monoatomic threads of carbon atoms are synthesized using the method of laser ablation in liquid. Irradiation of the distilled water solution of a carbonaceous precursor by a pulsed laser involves the phase transition in the precursor and results in the formation of the carbyne phase [12, 13]. The key peculiarity of the chosen mechanism of formation of carbon chains is the impact of metal NPs. NPs play several roles in the fabrication process. First, they catalyze the growth of linear carbon chains and simultaneously stabilize them [12, 13]. In the environment of a deterrent solution in the absence of metal NPs allows for assembling of linear carbon chains of the length of several atoms only. The presence of gold NPs allows stabilizing the monoatomic threads of the length of tens of atoms due to the formation of Au-C single electron bonds [13]. As a result, the carbyne chains capped with metal NPs at their ends become protected from folding and remain stable for a long time. For the samples obtained with use of this technique, the end-capped carbyne chains preserved their structural stability in a liquid media over several months.

Second, interestingly, gold NPs also provide an efficient tool for the manipulation of monoatomic chains. Clearly, charged NPs are sensitive to the external fields. In general, NPs can accumulate electric charges for various reasons [14]. In particular, they aquire an electric polarisation during the laser irradiation at the stage of the synthesis of carbyne wires. However, the charge distribution is stochastic in this case. In order to control charging of the NPs, we exploit the effect of Fermi level equalization between connected metal NPs. Being linked by the carbyne wires, two metal NPs exchange electrons until their electrochemical potentials equalize [15]. The amount of charge redistributed between NPs is expected to be dependent on their sizes. In particular, two identical NPs, which are brought into contact or connected by a thin conducting wire, would not exchange electrons. On the other hand, if the metal spherical NPs are of different diameters, the work functions of their electrons also differ [16]. Therefore, in order to bring the system into thermal equilibrium, the electrostatic energy of NPs must be changed to compensate the difference in their work functions. That is realized by transferring an electric charge between the NPs attached to different ends of the carbon chain. This charge exchange endows the NP-carbon wire complexes with the finite dipole moments and makes the linear chains of carbyne sensitive to the external electric field. This effect enables us to control the orientation of carbon chains at the stage of their surface deposition.

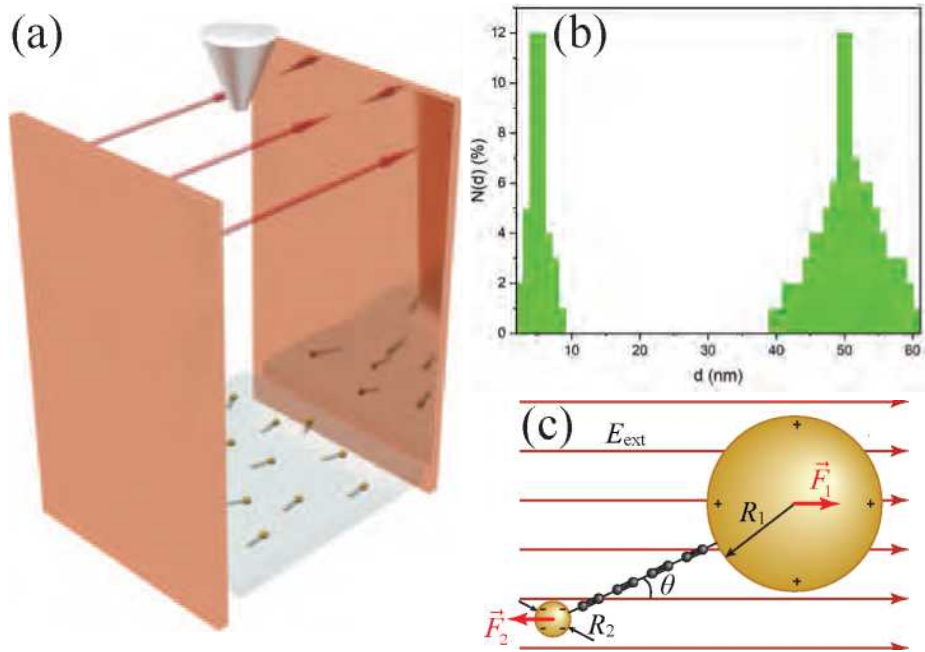

FIG. 1. (a) the scheme of the experimental setup used for the controlled deposition of carbon chains stabilised with gold NPs. The orientation of the electric field inside the capacitor if indicated by red arrows, (b) the histogram of the NP size distribution obtained by the dynamical laser scattering analysis. The percentage of gold NPs of the size d is shown in the vertical axis, (c) the carbon wire end-capped with gold NPs of different radii in the presence of the static electric field. The red arrows show the directions of Coulomb forces acting upon the gold NPs.

## Surface deposition of monoatomic carbon chains in the presence of static electric fields

In order to deposit aligned carbon chains on a substrate, we dissolve the carbonaceous precursor in the colloidal solution of NPs characterised by a bimodal size distribution, see Fig. 1b. The larger NPs were of the mean diameter of 50 nm and the smaller ones were of the mean diameter of about 5 nm. Figure 1a schematically illustrates the sputtering process we employed. After the laser ablation stage, the solution containing carbyne fraction passed under the pressure of 70 kPa between two parallel plane electrodes biased with a voltage ranging from 100 to 1000 V and spaced by 1-2 cm. In the presence of a stationary electric field of the magnitude up to $10^5$ V/m, carbon chains became oriented along the field lines during the deposition on a solid substrate.

To reveal the presence of an sp-carbon fraction in the deposited layer we study the Raman spectra of the samples, see Fig. 2a. The strong evidence of the presence of carbyne is provided by a pair of peaks at 1050 and 2150 cm$^{-1}$ related to single (C-C) and triple bonds (C≡C) of carbon, respectively. A broad maximum with a central frequency of 800 cm$^{-1}$ should be attributed to the mechanical distortion of single carbon bonds, responsible for the formation of the kinks on a straight linear chain (see Fig.

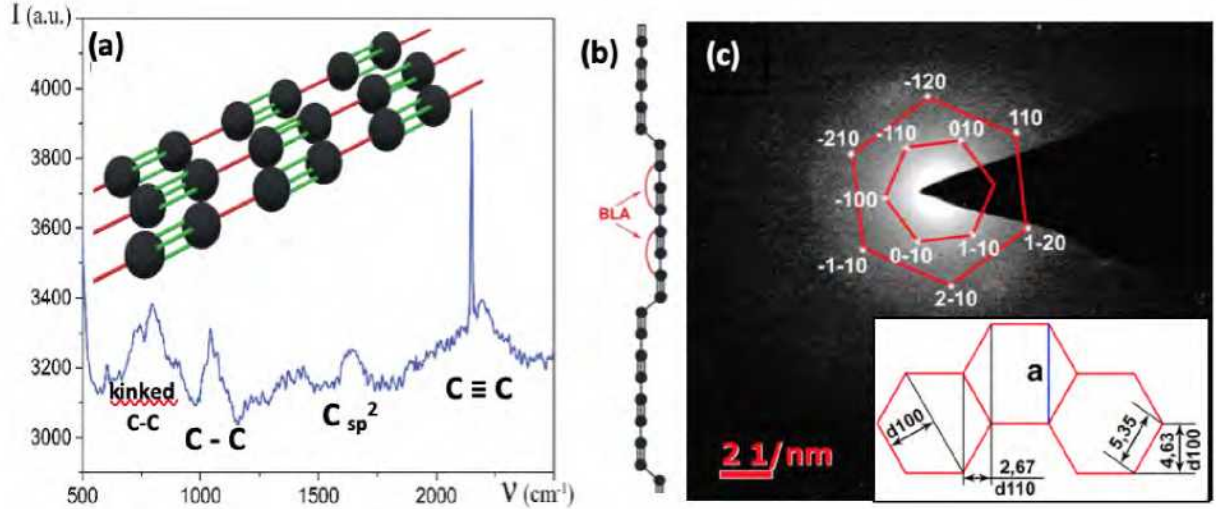

FIG. 2. Characterization of the deposited thin films: (a) the Raman spectrum; (b) The corner at the kinks is formed by two carbon atoms connected by a single C-C bond. The bond-length alternation is 2.57 Å for a polyyne chain; (c) the electron diffraction pattern of a thin deposited layer. The reflexes are labeled with the corresponding Miller indices. The insert shows the corresponding real-space structure of a carbyne crystal in the direction perpendicular to the chains. The parallel chains are hold together by the Van der Waals interaction.

2b), as well as to the contribution of the bonds connecting the chain with gold NPs. The presence of a peak near 1640 cm$^{-1}$ corresponds to the transformation of a certain amount of carbon to the sp$^2$ phase, which occurs due to the cross-linking between parallel linear chains, most probably. The appearance of a broad shoulder in the region 2100-2300 cm$^{-1}$ is due to the variation of the length of the polyyne bond caused by bending of the chains.

In order to proof the one-dimensional nature of the deposited substance we study the electron diffraction spectra of the samples in vacuum, see Fig. 2c. The distinct point-type reflexes indicate the presence of a 1D carbyne crystal that is characterised by a specific crystalline structure [17]. In particular, the observed hexagonal pattern shown in Fig. 2c corresponds to the bundle of linear carbon chains ordered in a hexagonal lattice having a zone axis [001]. The inset of Fig. 2c shows the structure extracted from the diffraction pattern, where each node of the lattice corresponds to a single carbon chain oriented perpendicular to the plane.

The inter-chain distances extracted from the diffraction patterns are d$_{100}$ = 4.63 Å and d$_{110}$ = 2.67 Å (see Fig. 2c), which are very close to the values for the hexagonal carbyne crystal obtained in [20]. This carbyne modification demonstrates lattice parameter of c = 9.26 Å, see Fig. 2b. The distance between neighboring parallel wires is 5.35 Å, that is almost twice larger than the bond-length alternation 2.57 Å in polyyne. Therefore the interaction between parallel chains is expected to be weaker than the interactions between neighboring carbon atoms in a chain. We conclude that the force responsible for the formation of a carbyne crystal is likely to be the Van der Walls force [7, 12, 18, 19].

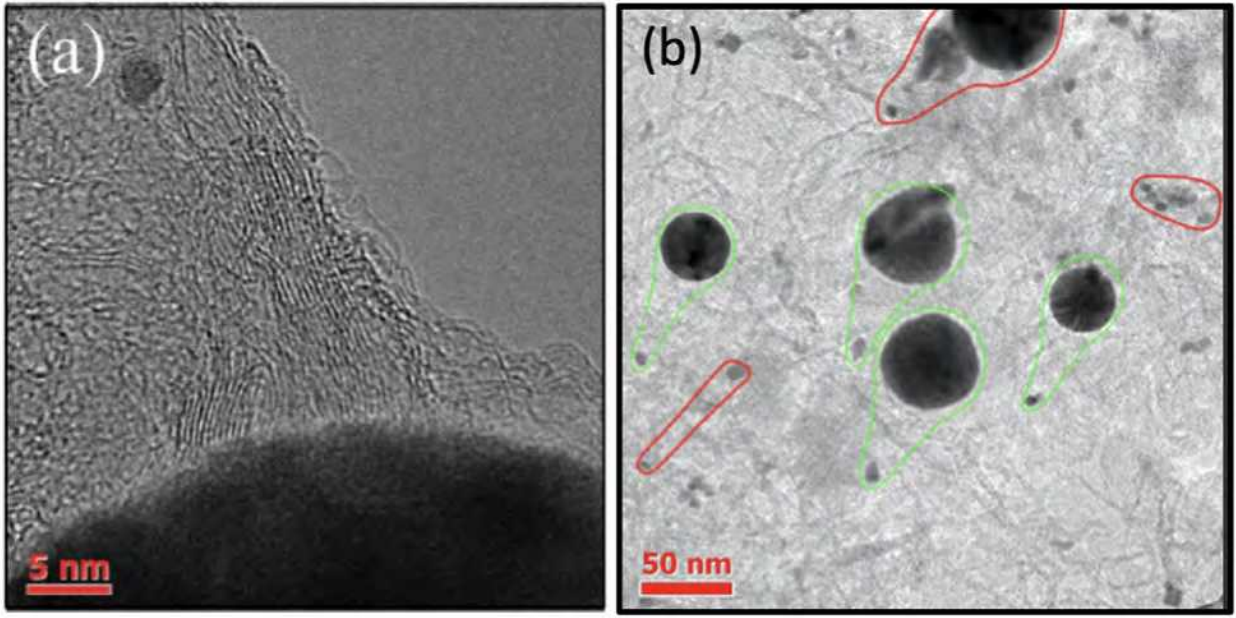

FIG. 3. Experimental technique employed for the deposition of monoatomic carbon chains end-capped with Au NPs in the presence of the static electric field. (a) TEM image of the carbyne wire bundles attached to two NPs. The dark region at the bottom corresponds to the larger NP of a nearly spherical shape with the radius ~ 25 nm. (b) - an ensemble of closely spaced nanodipoles oriented along the electric field direction (marked by green curves). The NP-carbyne complexes framed with red demonstrate no sensitivity to the electric field: they are randomly oriented.

For a direct visualization of the carbyne wires we used a high-resolution TEM technique. Fig. 3a depicts a bundle of parallel carbyne wires attached to two NPs of different radii: 2.5 and 25 nm, separated by about 50 nm. Ideally, a carbyne chain starts on the surface of one NP and ends on the surface of the other one. Indeed, in Fig. 3a one can see a bundle of about 20 threads (framed by a dashed line), which directly connects two NPs. However, there are also other threads which start growing at larger NP and form a closed loop attaching to the surface of the same NP. The orientation effect of the electric field is clearly seen on the lower-scale TEM image, Fig. 3b. In the presence of the field, the nanodipoles consisted of a larger and smaller NPs (framed by green curves) orient themselves along the field. The residual deflection angle between the nanodipole orientation and the electric field does not exceed few degrees in this particular case.

Note that the deposited layer contains also carbyne-NPs complexes consisting of the NPs of similar sizes (encircled by red curves in Fig. 3b). These complexes demonstrate a random orientation with respect to the field since they are not dipole-polarised.

The diagram showing the distribution of the angle of mutual orientation between the nanodipole and the field is shown in Fig. 4. In order to reproduce theoretically the dispersion in the orientation

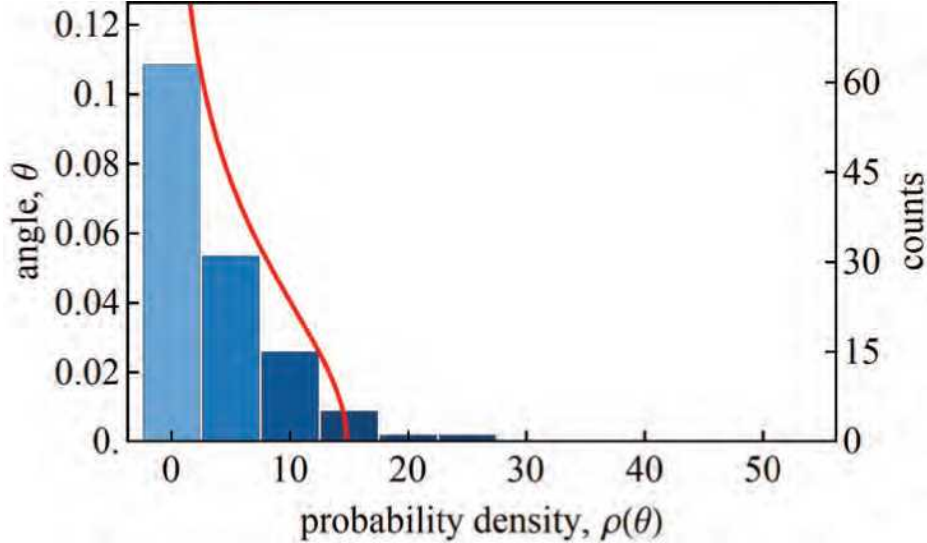

FIG. 4. Distribution of the angle between the deposited nanodipoles and the electric field. Bars indicate the results extracted from the TEM images of 116 individual nanodipoles. The dashed line corresponds to the probability density (9)

diagram we resort to the model describing the dynamics of a single carbyne- NPs nanodipole flying in the static electric field during the deposition process.

## DISCUSSION

### The kinetics of the field-assisted orientation of the carbyne chain end-capped with metal nanoparticles

The distribution of the electronic density between conducting NPs is governed by the Volta effect [21, 22]. Let us consider a couple of gold NPs connected by a thin conducting carbyne wire. We assume that before being connected, the NPs were electrically neutral. In this way, we disregard the random charging of NPs which might occur during the laser ablation process.

Each NP can be characterised by an electrochemical potential, $E < 0$, given by

$$E = -A - e\psi \tag{1}$$

where the electron work function A is the minimum energy required to remove an electron from inside the crystal to the point just outside its surface, and the electrostatic energy $-e\psi$ is the difference of electron potential energy at the crystal surface and that at the infinite distance from the NP. Here $\psi$ is an electrostatic potential of the conductor, $e$ is an elementary charge. It is important to note, that the work function of an NP differs from that of a bulk crystal, $A_{bulk}$. This difference is dependent on the size of the NP. For a neutral spherical NP of the radius $R$ surrounded by the dielectric medium with a permittivity $\epsilon$ the work function may be defined as [16, 23-25]

$$A(R) = A_{bulk} + \alpha \frac{e^2}{\epsilon R};$$ (2)

where $\alpha$ is a dimensionless constant. We shall estimate it as $\alpha = 3/8$, in agreement with the analytical theory [16] and the first-principles calculations [23]. The work functions of two NPs of different radii, namely, $R_1$ and $R_2$, differ by

$$\Delta A = \alpha \frac{e^2}{\epsilon} \frac{R_2 - R_1}{R_1 R_2}.$$ (3)

At thermal equilibrium, the electrochemical potential of a system of two NPs connected by a conducting wire must be constant. Once two NPs are put in contact, electrons start moving in the direction opposite to the gradient of the original electrochemical potential $E$, that is from the large NP to the small NP, see Fig. 1c. The steady state regime is achieved ones the work function mismatch is compensated by the modification of the electrostatic energy, see Eq. (1),

$$\Delta A = -e\Delta\psi.$$ (4)

We calculate the difference of electrochemical potentials $\Delta\psi$ between two conducting spheres with use of the capacitance matrix formalism [15, 26, 27]. For the sake of simplicity, we limit our consideration to the case of long inter-particle distances, where the length of the carbyne chain is comparable or exceeds the larger NP radius. Thus, for the Volta potential difference we take [27]

$$\Delta\psi = \frac{q}{\epsilon} \left( \frac{R_2 - R_1}{R_1 R_2} - \frac{2}{s} \right),$$ (5)

where s is the distance between the centers of NPs, and $q = q_2 = -q_1$ the transferred charge. Finally, combining (3), (4) and (5) we obtain the charge accumulated on the NP surface:

$$q = \frac{3}{8} \frac{R_2 - R_1}{R_1 R_2} \left( \frac{R_2 + R_1}{R_1 R_2} - \frac{2}{s} \right)^{-1} e.$$ (6)

As a result of the process described above, the most part of carbyne-NP complexes represent nano-dimensional dipoles. For $R_1 = 2.5$ nm, $R_2 = 25$ nm and $s \approx 50$ nm (this is the case shown in Fig. 3a) the dipole moment qs acquired because of the charge redistribution is about $10^3$ Debye.

The charge redistribution between NPs occurs at the stage of nanostructure fabrication in the colloid. The dipole-polarised carbyne-NP complexes are then launched from the nozzle towards the substrate. We assume that due to the pressure drop, the water is immediately desorbed from the NP surface. Therefore, the isolated nanodipoles, being subject to the static electric field $E_{ext}$, start rotating about their mass centers. The dynamics of this process obeys the simple kinetic equation of rotational motion

$$\partial_{tt}\theta = ML^{-1} - 2\gamma\partial_t\theta$$ (7)

for the angle $\theta$ between the wire and the electric field, see Fig. 1c. Here $M = sqE_{ext}sin\theta$ is a torque and $L = \frac{m_1 m_2}{m_1 + m_2}S^2 + \frac{2}{5}(m_1 R_1^2 + m_2 R_2^2)$ is a total momentum of inertia of NPs with masses $m_1$ and $m_2$, connected by the stiff massless wire. The last term in the right-hand side of Eq. (7) accounts for the dissipation of the rotational energy characterised by the rate $\gamma$ which we consider as a fitting parameter. The energy dissipation may be caused by the energy transfer to the vibrations of the carbyne wire or by the interaction of the nanodipole with water droplets. Expressing the masses of the NPs through their radii, $m_{1,2} = \frac{4\pi}{3}\rho_{Au}R_{1,2}^3$, and taking into account that $R_2 \gg R_1$, we obtain

$$\partial_{tt}\theta + 2\gamma\partial_t\theta + \omega_0^2 sin\theta = 0 \qquad (8)$$

where $\omega_0 \approx \sqrt{\frac{45e}{64\pi\rho_{Au}}\frac{s}{R_2^5}E_{ext}}$. Note that Eq. (8) describes the dynamics of a simple gravitational pendulum, so $\omega_0$ defines the oscillation frequency of the nanodipole provided that the initial angular displacement $\theta_0$ is small. Although the dipoles are randomly oriented at the exit of the nozzle, the narrow resulting distribution of their orientation angle is observed in the experiment, see Fig. 4. This indicates that the dumping is efficient. Hereafter, we neglect the nonlinear terms in Eq. (8), tacking $sin\theta \approx \theta$. For the radii of NPs characteristic of our experiments and the electric field $E_{ext} = 10^5\ V/m$, we obtain the oscillation period $T_0 = 2\pi/\omega_0$ of the order of a few microseconds. The period scales as $T_0 \propto d^{-1/2}$, where d is the carbyne wire length. In order to examine the ability of the field to align the nanodipoles, we compare this value with the time of flight of the NPs in the presence of the external electric field during the sputtering process, $\tau = l/v$. Here $l$ is a distance between the nozzle and the substrate, $v \approx \sqrt{2\Delta p/\rho_{sol}}$ is the initial mean particle velocity with $\Delta p$ being the excess pressure and $\rho_{sol}$ being the density of the sputtered solution, which we take equal to the density of water. In our experiments, the time of flight is about $10^{-2}$ s. It exceeds the period of the nanodipole oscillations $T_0$ by more than three orders of magnitude. We conclude that the capping of the carbyne wires with metal nanoparticle of different sizes makes them very sensitive to the external electric field.

In order to estimate the angular distribution of deposited carbon chains, we assume that the dispersion of the time of flight of different nanodipoles exceeds the average period of oscillations $T_0$. Therefore, for a given nanodipole, which starts oscillating from the angle $\theta_0$, the probability of being oriented in the direction $\theta$ with respect to the electric field is $\sigma(\theta, \theta_0) = 2|\partial_t\theta(t, \theta_0)|^{-1}/T_0$. Then, assuming that the damping rate $\gamma$ is much smaller than the oscillation frequency $\omega_0$, that is consistent with our experimental findings, we obtain: $\sigma(\theta, \theta_0) = \frac{exp(\gamma\tau)}{\sqrt{\theta_0^2 - \theta^2 exp(2\gamma\tau)}}$, where $m$ is the average time of flight. Tacking into account that the nanodipoles are randomly oriented at the initial moment of time (at the exit of the nozzle), we obtain their angular distribution on the substrate:

$$\rho(\theta) = \frac{1}{2\pi}\int_{|\theta_0|<|\theta|}\sigma(\theta, \theta_0)d\theta_0 = \frac{exp(\gamma\tau)}{2\pi^2}Re\left[ln\left(\frac{\pi + \sqrt{\pi^2 - \theta^2 exp(2\gamma\tau)}}{\pi - \sqrt{\pi^2 - \theta^2 exp(2\gamma\tau)}}\right)\right]. \qquad (9)$$

This distribution is shown in Fig. 4 by a red line. The best fit to the experimental data was achieved with the damping rate $\gamma = 2.5\tau^{-1}$. The small mismatch between the experimental data and the predictions of a simple kinetic model might result from the stochastic fluctuations of the total amount of charge accumulated on the NP surfaces. It is noteworthy that in spite of these fluctuations, a systematic charge redistribution occurs between a pair of NPs with different diameters. Therefore, any difference in the NP charges would trigger the precession of NP-carbyne complexes in the electric field. This precession damped due to the interactions with the environment is a key mechanism of the observed ordering of carbyne wires.

## CONCLUSION

In conclusion, we demonstrated an efficient approach for the fabrication of spatially oriented ensembles of monoatomic carbon chains stabilised by gold NPs and deposited on the substrate. The fabrication technique that we employed consists in the sputtering of the colloidal solution of NP-carbyne complexes in the presence of the stationary electric field. We use the mixture of two sorts of Au NPs whose diameters differ by an order of magnitude. In this way, we induce the charge redistribution between the NPs of different sizes connected by conducting monoatomic carbon wires. The resulting dipole polarisation makes the carbyne-NP complexes very sensitive to the external electric field which tends to order all the nanodipoles along its direction. The orientation diagram of carbyne-NP complexes on a substrate evidences their ordering along the applied electric field direction.

These results demonstrate that the fabrication of monoatomic carbon wires end-capped with metal nanoparticles provides a versatile tool for harnessing of the fascinating properties of carbyne. The fabrication of nanoribbons consisting of parallel monoatomic carbon chains is a promising step towards the realisation of nano-electronic networks based on carbyne- gold complexes.

## METHODS

We have used the method of laser fragmentation of colloidal carbon systems for realisation of stabilised carbyne chains. The method is described in detail in [11]. To create carbon-gold bonds we additionally illuminated the solution by nanosecond laser pulses generated by an Ytterbium (Yb) fiber laser having the central wave length of 1.06 ^m, the pulse duration of 100 ns, the repetition rate of 20 kHz and the pulse energy of up to 1 J. The time between subsequent pulses was about 1 s. The sizes of gold NPs have been controlled by the dynamic laser scattering device Horiba SZ100.

The Raman spectra were measured by using the Senterra spectrometer, made by Bruker company. The pump laser wavelength is of 532 nm at the power level of 40 mW, the radiation was focused through a 50-fold microlens, the spectra were collected in the confocal microscope configuration and averaged over 10 measurements. The accumulation time of each measurement was 60 seconds.

For the detailed study of the orientation distribution of nanodipoles, we have performed the high resolution transmission electron microscopy and X-ray diffraction studies using FEI Titan³ with a spatial resolution of up to 2 Å. Processing of TEM-images and diffraction patterns was conducted with the opened database package Image J 1.52 a.

## ACKNOWLEDGMENTS

The work of S.K., I.C. and A.KA. is supported by the Westlake University (Project No. 041020100118). This work was also partially supported by RFBR grants 17-32-50171, 18-3220006 and 19-32-90085. The synthesis and deposition of carbynes have been performed at the Vladimir State University. Raman spectra was measured at the Center for Optical and Laser Materials Research, Research Park, St. Petersburg State University. TEM measurements have been performed in the "System for microscopy and analysis" LLC, Moscow.

## AUTHOR CONTRIBUTIONS

S.K. has conceived the work and analysed experimental data;
I.C. contributed to the theoretical model and reviewed the manuscript;
A.K. contributed to the synthesis of monoatomic carbon chains and realized HR TEM microscopy study;
A.O. performed the laser experiments and Raman spectrum collection;
V.S. realized the carbyne deposition on a substrate;
I.S. analyzed the diffraction data;
A.KA. contributed to the interpretation of the results and has coordinated the collaborative work.

# Suplementary Material to Electric field assisted alignment of monoatomic carbon chains

Stella Kutrovskaya, Igor Chestnov, Anton Osipov, Vlad
Samyshkin, Irina Sapegina, Alexey Kavokin, and Alexey Kucherik

Here we describe the approach that allowed us to collect the data of the orientation of deposited carbyne - gold nanoparticles (NPs) complexes by TEM scaling. Figure 1 illustrates the sputtering set-up geometry (a) with reference points marking the electric field orientation on the surface of the sample. (b) shows the sample fixed on the TEM holder as well at the frame chosen for the orientation measurements, (c) shows an example of the TEM image of the deposited nanostructures with the greed line showing the orientation of a specific structure, red line showing the electric field orientation. Figure 2 shows the magnified TEM image of an individual deposited nanostructure composed by a carbyne thread connecting two gold NPs of different sizes. The green line connects the centers of NPs. The inset shows the angle between the green line and the red line that corresponds to the electric field orientation. Figure 3 shows TEM images of a reference sample where carbyne-gold nanostructures were deposited on a substrate in the absence of external electric field. One can clearly see that no any alignment or orientation of deposited structures is observed in this case.

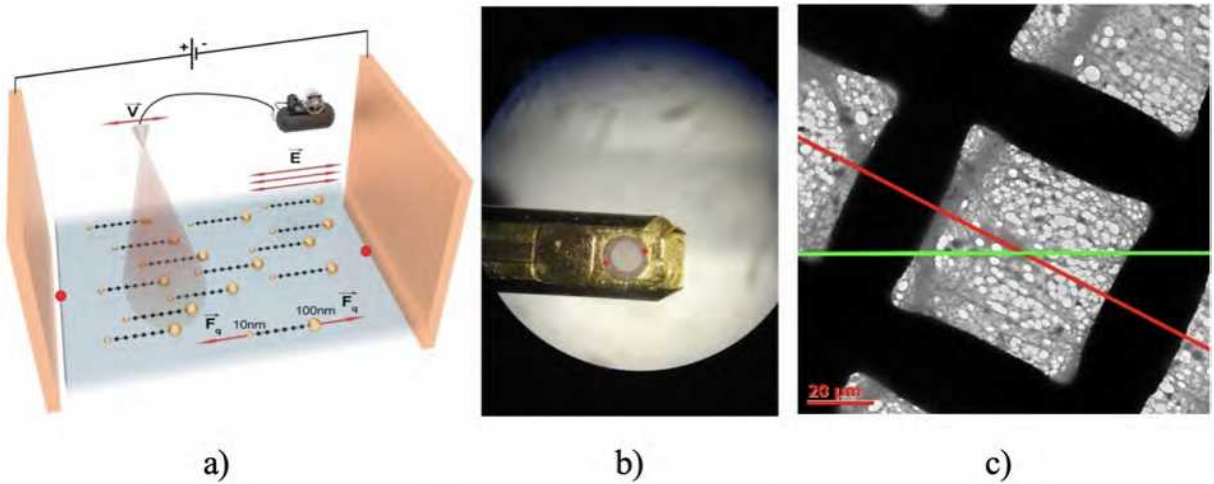

a)             b)             c)

FIG. 1. The method of measurement of the angle of orientation of the deposited nanodipoles with respect to the electric field orientation during the sputtering process. (a) The scheme of the sputtering setup showing the reference points that mark the direction corresponding to the electric field orientation of the surface of the substrate. (b) The view of a sample fixed on a TEM holder, where two red points set up the reference frame. (c) The green line shows the orientation of one of the deposited nanostructures, while the red line shows the orientation of the electric field. The angle between red and green lines is measured for each of studied nanostructures in order to obtain the statistical distribution.

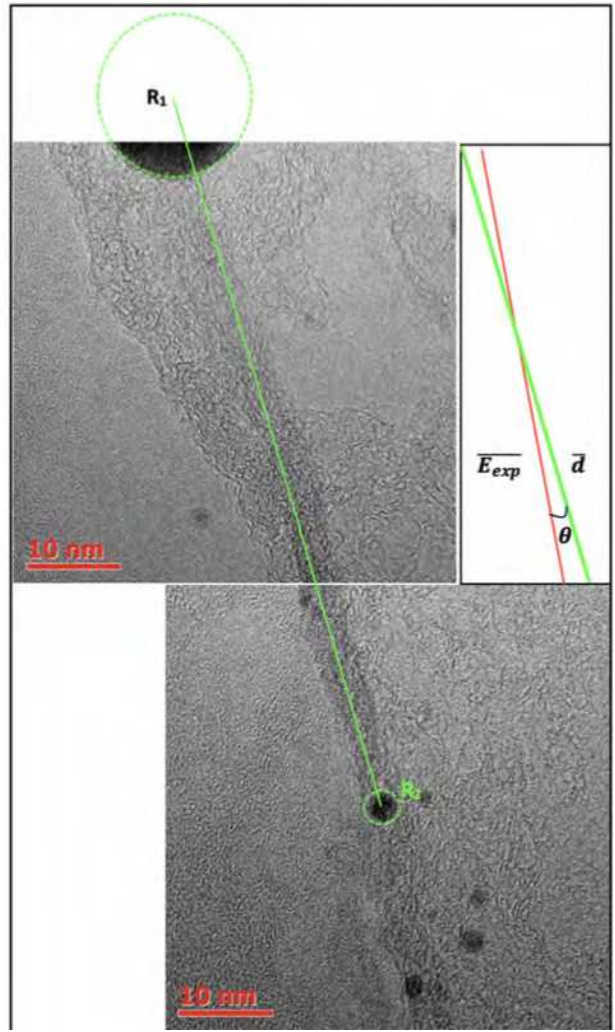

FIG. 2. The angle 9 between the orientation of the deposited carbyne-gold nanostructure and the electric field is extracted from the analyses of high-resolution TEM images. The green line connecting the centers of NPs on the TEM image is superposed to the red line showing the electric field orientation in the inset. The values of 9 angles characterising the orientation of all studied nanostructures are stored in the computer memory and used to produce the experimental distribution function.

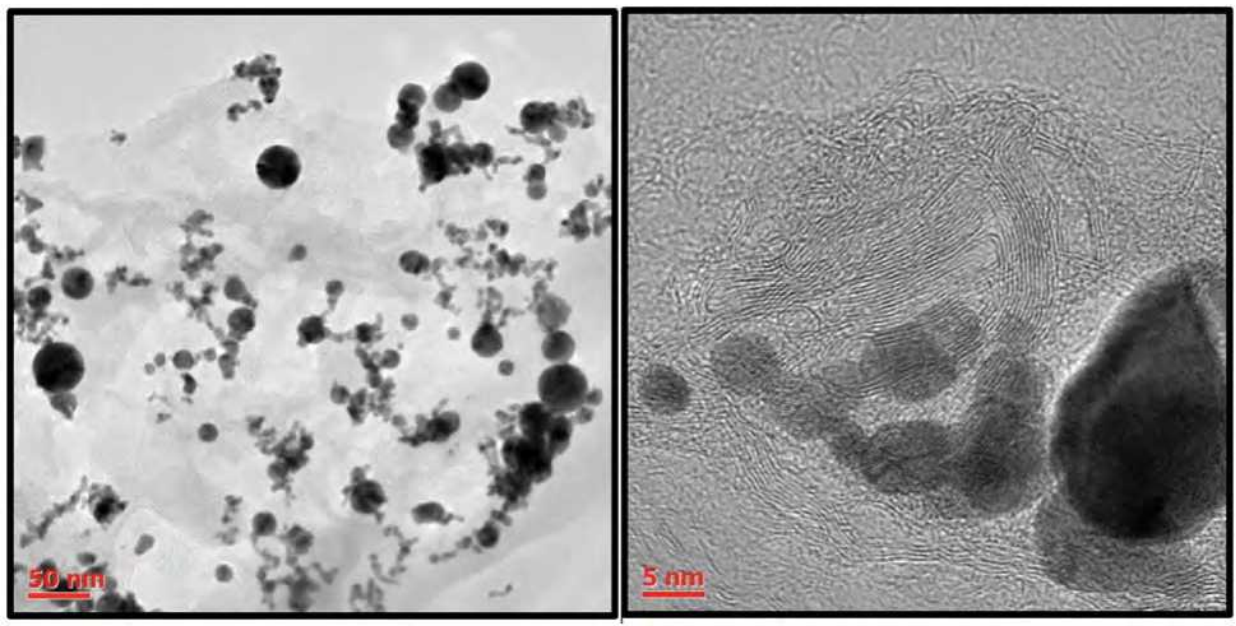

FIG. 3. Experimental results on the reference sample where the deposition of the monoatomic carbon chains end-capped with Au NPs has been done by sputtering in the absence of the static electric field. Left and right panels show lower and higher spatial resolution images, respectively.